\documentclass[10pt]{article}
\usepackage{amssymb}
\usepackage{amsmath}
\usepackage[english]{babel}
\usepackage{doublespace}
\usepackage{stmaryrd}
\renewcommand{\baselinestretch}{1.2}
\normalsize
\makeatletter
\def\@begintheorem#1#2{\trivlist%
 \item[\hskip \labelsep{\sffamily\bfseries #2\ #1}]\itshape}
\newtheorem{_teo}{Theorem}[section]
\newenvironment{teo}{\begin{spacing}{1.2} \begin{_teo}}{\end{_teo}\end{spacing}}
\newtheorem{_defi}[_teo]{Definition}
\newenvironment{defi}{\begin{spacing}{1.2} \begin{_defi}\rm}{\end{_defi}\end{spacing}}
\newtheorem{_rem}[_teo]{Remark}
\newenvironment{rem}{\begin{spacing}{1.2} \begin{_rem}\rm}{\end{_rem}\end{spacing}}
\newtheorem{_cor}[_teo]{Corollary}
\newenvironment{cor}{\begin{spacing}{1.2} \begin{_cor}}{\end{_cor}\end{spacing}}
\newtheorem{_lem}[_teo]{Lemma}
\newenvironment{lem}{\begin{spacing}{1.2} \begin{_lem}}{\end{_lem}\end{spacing}}
\newtheorem{_pro}[_teo]{Proposition}
\newenvironment{pro}{\begin{spacing}{1.2} \begin{_pro}}{\end{_pro}\end{spacing}}
\newenvironment{prop}{\begin{spacing}{1.2} \begin{_pro}}{\end{_pro}\end{spacing}}
\newtheorem{_eje}[_teo]{Example}

\newtheorem{_claim}[_teo]{Claim}

\makeatother

\newenvironment{beweis}{\noindent{\em Proof.}}{\hfill $\rule{2mm}{2mm}$
\vspace{3mm}\par}
\newenvironment{prf}{\noindent{\em Proof.}}{\hfill $\rule{2mm}{2mm}$
\vspace{3mm}\par}

\DeclareMathAlphabet{\Ma}{U}{msa}{m}{n}
\DeclareMathAlphabet{\Mb}{U}{msb}{m}{n}
\DeclareMathAlphabet{\Meuf}{U}{euf}{m}{n}

\def\got#1{\Meuf{#1}}

\DeclareSymbolFont{ASMa}{U}{msa}{m}{n}
\DeclareSymbolFont{ASMb}{U}{msb}{m}{n}
\DeclareMathSymbol{\hrist}{\mathord}{ASMa}{"16}
\DeclareMathSymbol{\varkappa}{\mathalpha}{ASMb}{"7B}
\DeclareMathSymbol{\CrPr}{\mathord}{ASMb}{"6F}

\newfont{\EinsFont}{cmr7 scaled 1070}

\def\restriction{{\mathchoice{
 \mbox{\unitlength1cm\begin{picture}(.2,.4)%
  \bezier{5}(.07,.3)(.1,.27)(.13,.24)%
  \put(.07,.35){\line(0,-1){.5}}\end{picture}}}{
 \mbox{\unitlength1cm\begin{picture}(.2,.4)%
  \bezier{5}(.07,.3)(.1,.27)(.13,.24)%
  \put(.07,.35){\line(0,-1){.5}}\end{picture}}}{
  \hrist}{\hrist}}}

  \def\al #1.{{\mathcal{#1}}}
  \def\ot #1.{{\got{#1}}}

  \def\ccr #1,#2.{\overline{\Delta(#1,\,#2)}}

  \def\b #1.{{\bf #1}}
  \def\cross#1.{\mathrel{\mathop{\times}\limits_{#1}}}
  
  \def\C{\Mb{C}}
  \def\F{\Mb{F}}
  \def\N{\Mb{N}}
  
  \def\R{\Mb{R}}
  
  \def\T{\Mb{T}}
  
  \def\Cn{{\bf V}}

  \def\wt{\widetilde}

\def\cross #1.{\mathrel{\raise 3pt\hbox{$\mathop\times\limits_{#1}$}}}
\def\maprightu #1;{\mathrel{\smash{\mathop{\longrightarrow}\limits^{#1}}}}
\def\maprightb #1;#2.{\mathrel{\smash{\mathop{\longrightarrow}\limits_{#2}^{#1}}}}
\def\s #1.{_{\smash{\lower2pt\hbox{\mathsurround=0pt $\scriptstyle #1$}}\mathsurround=5pt}}
\def\cl #1.{{\cal #1}} \def\al #1.{{\cal #1}}
\def\ker{{\rm Ker}\,}
\def\aut{{\rm Aut}\,}
\def\gau{{\rm Gau}\,}

\def\f #1,#2.{\mathsurround=0pt \hbox{${#1\over #2}$}\mathsurround=5pt}
\def\hlf{{\f 1,2.}}
\def\ol#1.{\overline{#1}}
\def\q{{\bf q}}
\def\CAR{{{\rm CAR}(L^2(E))}}
\def\chop{\hfill\break}

\def\fl #1,#2.{{\mathord{\rm Fl}_{#1}(#2)}}
\def\GL{\mathord{\rm GL}}

\def\XP#1!{\renewcommand{\baselinestretch}{.7}\marginpar{{\footnotesize
$\leftarrow$#1}\hfil}
\renewcommand{\baselinestretch}{1.5}}
\def\XB{\marginpar{
{\footnotesize\bf Change~starts-----}\lower 11pt\hbox{\mathsurround=0pt$
\!\!\displaystyle{
\Bigg\downarrow}$\mathsurround=3pt}}}
\def\XE{\marginpar{{\footnotesize\bf Change~ends-----}\raise 10pt\hbox{\mathsurround=0pt$
\!\!\displaystyle{
\Bigg\downarrow}$\mathsurround=3pt}}}

\newcommand{\id}{\mathop{{\rm id}}\nolimits}

\newcommand{\Diff}{\mathop{{\rm Diff}}\nolimits}
\newcommand{\Gau}{\mathop{{\rm Gau}}\nolimits}
\newcommand{\Aut}{\mathop{{\rm Aut}}\nolimits}

\newcommand{\subeq}{\subseteq}

\renewcommand{\:}{\colon}
\newcommand{\oline}{\overline}

\newcommand{\im}{\mathop{{\rm im}}\nolimits}

\newcommand{\U}{\mathop{\rm U{}}\nolimits}
\newcommand{\Car}{\mathop{{\rm CAR}}\nolimits}
\newcommand{\Gicar}{\mathop{{\rm GICAR}}\nolimits}

\newcommand{\Rarrow}{\Rightarrow}

\newcommand{\3}{\ss{}}

\newcommand{\cA}{\mathcal{A}}
\newcommand{\cB}{\mathcal{B}}
\newcommand{\cF}{\mathcal{F}}
\newcommand{\cN}{\mathcal{N}}
\newcommand{\cH}{\mathcal{H}}
\newcommand{\cP}{\mathcal{P}}
\newcommand{\cU}{\mathcal{U}}
\newcommand{\cL}{\mathcal{L}}
\newcommand{\cK}{\mathcal{K}}
\newcommand{\la}{\langle}
\newcommand{\ra}{\rangle}
\newcommand{\1}{\mathbf{1}}
\newcommand{\eps}{\varepsilon}
\newcommand{\Spann}{\mathop{{\rm span}}\nolimits}
\newcommand{\bi}{\mathbf{i}}
\newcommand{\bj}{\mathbf{j}}
\newcommand{\bv}{\mathbf{v}}

\title{\bf Localization via Automorphisms of the CARs.
Local gauge invariance}

\author{
 {\sc Hendrik Grundling}                                            \\[1mm]
 {\footnotesize Department of Mathematics,}                         \\
 {\footnotesize University of New South Wales,}                      \\
 {\footnotesize Sydney, NSW 2052, Australia.}                             \\
 {\footnotesize hendrik@maths.unsw.edu.au}                  \\
 {\footnotesize FAX: +61-2-93857123}   \\
\and
  {\sc Karl--Hermann Neeb}              \\[1mm]
 {\footnotesize Fachbereich Mathematik,}                           \\
 {\footnotesize Technische Universit\"at Darmstadt,}    \\
 {\footnotesize  Schlossgartenstrasse 7,}                      \\
 {\footnotesize D--64289 Darmstadt Germany.}                        \\
 {\footnotesize neeb@mathematik.tu-darmstadt.de}}

\begin{document}
\maketitle

\begin{abstract}

\noindent
The classical matter fields are sections of a
vector bundle $E$ with base manifold $M,$ and the
space $L^2(E)$ of square integrable matter fields
w.r.t. a locally Lebesgue measure on $M,$
has an important
module action of $C_b^\infty(M)$ on it. This module action
defines restriction maps and encodes the
local structure of the classical fields.
For the quantum context, we show that this module action defines
an automorphism group on the algebra of the canonical anticommutation relations,
$\CAR,$ with which we can perform the analogous localization.
That is, the net structure of the $\CAR$ w.r.t. appropriate subsets of $M$ can be
obtained simply from the invariance algebras of appropriate subgroups.
We also identify the quantum analogues of restriction maps,
and as a corollary, we  prove a well--known ``folk theorem,'' that
the $\CAR$ contains only trivial gauge invariant observables w.r.t. a
{\it local} gauge group acting on $E.$
\end{abstract}

\section{Introduction}

\noindent
Classically, a matter field $\Psi$ on spacetime $M$ is a smooth section of an appropriate smooth vector
bundle $\q_E:E\to M$
with typical fiber being a finite dimensional vector space  $\Cn$ over $\F\in\{\R,\C\}.$
The space $\Gamma(E)$ of smooth sections of $E$ is a module for the action of pointwise multiplication
with $C^\infty(M):=C^\infty(M,\F),$ and this module action encodes the local structure
of the fields. In particular, for an open set $U\subset M$ the submodule
carried by $U,$ i.e. $\Gamma_U
:={\{c\in\Gamma(E)\,\mid\, c(U^c)=0\}},$ is
\[
\Gamma_U=\overline
{{\rm Span}\{f\cdot c\,\mid\,c\in\Gamma(E),\; f\in C^\infty(M),\;
f(U^c)=0\}}.
\]
In a quantum field theory on the other hand, one is  given a
$C^*$-algebra $\al F.$ (the ``field algebra'')
in which spacetime locality is specified by the following:
\begin{itemize}
\item[(L1)] There is  directed set  $\Upsilon$ of open relatively
compact subsets of $M$,
  partially ordered by set inclusion, such that
 $M=\bigcup\{W\mid W\in\Upsilon\}$ and
 $\xi(W)\in\Upsilon$ for all $W\in\Upsilon$ and $\xi\in\Diff M$.
 Moreover each $\xi\in\Diff M$ is uniquely determined by its action on $\Upsilon.$
\item[(L2)] There is an injective map $\al A.$ from $\Upsilon$ to the unital C*--subalgebras of $\al F.$  which is
order preserving, i.e. $\al A.(W_1)\subseteq\al A.(W_2)$ if $W_1\subseteq W_2.$
\end{itemize}
Further relativistic structures are usually also given (cf.~\cite{bHaag92,Haag64}), but here we will
not be concerned with those. There is usually no counterpart of the classical
$C^\infty(M)\hbox{-module}$ action, and given an $A\in\al F.,$  no
restriction map of $A$ to $W,$ producing an element of $\al A.(W),$ is specified.

Here we will be concerned with this issue;- in particular, we will show that for
the C*-algebra of the canonical anticommutation relations, built
upon the classical matter fields,
that the classical module structure of the algebra
$C_b^\infty(M)$  (resp. $L_\C^\infty(M))$ on $L^2(E)$,
defines an automorphic action
$\alpha \:\U\big( L_\C^\infty(M,\mu)\big)\to\aut\CAR,$ such that
\begin{itemize}
\item
 each $\al A.(W)$ is  the fixed point algebra in $\CAR$ of
the automorphism group $\alpha\s \U(L_\C^\infty(W^c)).,$
hence $\alpha$  contains the
locality information in this quantum context.
In this last expression we used the
natural identification of $\U(L_\C^\infty(W^c))$
with the unitaries in $\U\big( L_\C^\infty(M)\big)$
which are $1$ on $W,$ i.e., with $\chi_W+\U(L_\C^\infty(W^c)).$
\item
We will obtain
a conditional expectation $\nu_W:\CAR\to \al A.(W)$ which is exactly the
quantum restriction map of observables to $\al A.(W).$
\item
As a corollary, we are  able to prove the well--known ``folk theorem,'' that
the $\CAR$ contains only trivial gauge invariant observables w.r.t. a
{\it local} gauge group acting on~$E.$
\end{itemize}

\section{Commutative von Neumann Algebras and Automorphisms
of the CAR}

In this section, we will first prove in full generality the appropriate
properties of the $\Car(\cH)$ which we will need, and in the subsequent sections
will apply these.

\begin{defi}
Let $\cH$ be a complex Hilbert space (not necessarily separable).
A {\it CAR-algebra} of $\cH$ is a $C^*$-algebra
$\Car(\cH)$, together with a continuous antilinear map
$a \: \cH \to \Car(\cH)$ whose image generates $\Car(\cH)$
as a $C^*$-algebra and which satisfies the
{\it canonical anticommutation relations}
\begin{equation}
  \label{eq:car}
\{a(f), a(g)^*\}  = \la f, g \ra \1
\quad \hbox{ and } \quad  \{a(f),a(g)\} = 0
\quad \mbox{ for } \quad f, g \in \cH
\end{equation}
where we write $\{A,B\} := AB + BA$ for the anticommutator
of two operators. This determines
$\Car(\cH)$ up to natural isomorphism (\cite[Thm.~5.2.5]{BR97}), in particular,
it is a simple $C^*$-algebra. In view of the naturality,
there is an automorphic action
$\alpha: \U(\cH)\to\aut{\rm CAR}(\al H.)$
given by $$
{\alpha\s U.\big(a(f)\big)}:=a(Uf) \quad \mbox{ for } \quad U\in \U(\al H.),\; f\in\al H..$$
The pointwise continuity of the action (where $\U(\al H.)$ has the strong operator topology)
is immediate from the continuity and the
$\U(\al H.)\hbox{-equivariance}$ of the map $a \: \cH \to \Car(\cH).$
\end{defi}

The first main theorem which we want to prove in this section is the following:
\chop\chop
\noindent{\bf Theorem.}$\;$ {\it
Let $\cH$ be complex Hilbert space (not necessarily separable) and let
${\cal N} \subeq B(\cH)$ be a non-atomic commutative von Neumann algebra.
Then the fixed point algebra of the action $\alpha:\U(\cl N.)\to\aut{\rm CAR}(\al H.)$
is trivial, i.e. $\Car(\cH)^{\U(\cl N.)} = \C \1$.}
\chop\chop
The proof for this is long, and requires some preparatory results.

\begin{prop}
\label{NonatomP}
 For a commutative
von Neumann algebra $\cN \subeq B(\cH)$, the
following are equivalent:
\begin{description}
\item[\rm(a)] $\cN$ is non-atomic, i.e. $\cN$ does not contain any
minimal non-zero projection.
\item[\rm(b)] There exists a weakly continuous curve
$\gamma \: [0,1] \to P_\cN := \{ p  \in \cN \mid p^* =p = p^2\}$
with $\gamma(0) = \1$ and $\gamma(1) = \1$, i.e.
$P_\cN$ is arcwise connected in the weak topology.
\item[\rm(c)] For each $v \in \cH$ and each $\eps > 0$ there exists a
finite set of mutually orthogonal
projections $P_1, \ldots, P_r \in \cN$ with $\sum_{j = 1}^r P_j = \1$
and $\max_j \|P_jv\| < \eps$.
\end{description}
\end{prop}
\begin{prf}
(a) $\Rarrow$ (b), (c): Since every non-atomic
commutative von Neumann algebra $\cN$ is an $\ell^\infty$-direct
sum of algebras of the form
$L^\infty(X,\mu)$, where $(X,\mu)$ is a non-atomic measure space,
it suffices to prove the assertion for such an algebra $L^\infty(X,\mu)$.
Now \cite[Lemma~2.5]{GN01} implies the existence of an increasing
family $(X_t)_{0 \leq t \leq 1}$ of measurable subsets
of $X$ with $\mu(X_t) = t$.
We then put $\gamma(t) := \chi\s{X_t}.$ (the characteristic function).
To verify that $\gamma$ is $\sigma(\cN,\cN_*)$-continuous,
we have to show  that for each $f \in L^1(X,\mu) \cong L^\infty(X,\mu)_*$,
the curve $t \mapsto \la f, \gamma_t \ra = \int_{X_t} f\, d\mu$
is continuous, which follows from Lebesgue's Theorem on
Dominated Convergence.

To verify (c), let $v \in \cH$ and observe that
$f_v(A) := \la Av, v\ra$ defines an element of $\cN_*$
with $\|Pv\|^2 = f_v(P)$ for each projection $P \in P_\cN$.
Since $f_v \circ \gamma \: [0,1] \to \R$ is continuous,
it is uniformly continuous, and there exists an $n \in \N$
with
$$ \Big|f_v(\gamma\Big(\frac{k}{n}\Big))
- f_v(\gamma\Big(\frac{k-1}{n}\Big))\Big| < \eps
\quad \mbox{ for } \quad k =1,\ldots, n. $$
For the projections
$$P_k := \gamma\Big(\frac{k}{n}\Big)\Big
(\1 - \gamma\Big(\frac{k-1}{n}\Big)\Big)
= \gamma\Big(\frac{k}{n}\Big)- \gamma\Big(\frac{k-1}{n}\Big) $$
we then have
$$\|P_k v\|^2 = f_v(P_k) < \eps
\quad \mbox{ and } \quad
P_1 + P_2  + \cdots + P_n = \1. $$

(b) $\Rarrow$ (a): If $\cN$ is not non-atomic and $P$ is a minimal
non-zero projection, then the curve
$P \gamma(t)$ in $P_\cN$ is weakly continuous from
$0$ to $P$, which contradicts the minimality of $P$.

(c) $\Rarrow$ (a): If $\cN$ is not non-atomic and $P$ is a minimal
projection, then we pick a unit vector $v \in \im(P)$.
The minimality of $P$ implies that either $P_jP = 0$
or $P_jP = P$, so that either $P_j v = 0$ or $P_j v = v$.
This leads to $\max_j \|P_j v \| = \|v\|=1$ which contradicts (c).
\end{prf}
In the case that $\cH$ is separable, this proposition follows
from the fact that non-atomic maximal abelian
von Neumann algebras are isomorphic to $L^\infty([0,1])$
(cf. \cite[Thm.~9.4.1]{KR2}).
\begin{lem} \label{lem:finpro}
Let $\cP := \{ P_1, \ldots, P_r\}$ be a
set of nonzero mutually orthogonal projectors with
$\sum\limits_{j=1}^rP_j=\1$ (called a partition). Let $T := \prod_{j = 1}^r e^{i\R P_j}
\cong \T^r \subeq \U(\cH)$ be the corresponding unitary group and let
$$ \nu\s{\cP}. \: \Car(\cH) \to \Car(\cH)^T, \quad
\nu\s{\cP}.(A) := \int_T \alpha_t(A)\, d\mu_T(t) $$
be the fixed point projection, where $\mu_T$ is the normalized
Haar measure on the torus $T$. Let
$$A(f_1,\ldots,f_n;g_1,\ldots,g_n)
:=a(f_1)^*\cdots a(f_n)^*a(g_1)\cdots a(g_n)
\quad \mbox{ for } \quad f_i, g_j \in \cH $$
and
\[ \cB_n := \big\{A(f_1,\ldots,f_n;g_1,\ldots,g_n)
\,\mid\,f_i,\,g_i\in\al H.\big\},\;\;\al B._0:=\C\1. \]
Then there exists for each
$A \in \Spann\big(\bigcup_{n = 0}^\infty \cB_n\big) \subeq \Car(\cH)$
a constant $C_A$, independent of $\cP$,
such that for each $f \in \cH$ we have the estimate
$$ \| [\nu_{\cP}(A), a(f)]\| \leq C_A \max_\ell \| P_\ell f\|. $$
\end{lem}

If $\cP = \{\1\}$, then $T = \T \1$ and
$\Car(\cH)^T = \Gicar(\cH)$ is the well-known GICAR (``gauge invariant CAR'')
which is the closure of $\Spann\big(\bigcup_{n = 0}^\infty \cB_n\big).$
For any other partition $\cP,$ the algebra $\Gicar(\cH)$
contains $\Car(\cH)^T$ for the corresponding $T.$

\begin{prf} First we observe that for
$\bi = (i_1, \ldots, i_n), \bj = (j_1, \ldots, j_n)
\in \{1,\ldots, r\}^n$, the element
$$ A(P_{i_1}f_1,\ldots,P_{i_n}f_n; P_{j_1}g_1,\ldots,P_{j_n}g_n)
\in \Car(\cH) $$
is an eigenvector of $\alpha_t,$ $t\in T,$ corresponding to the character
$$ t = (t_1,\ldots, t_r) \mapsto \prod_{k = 1}^n t_{i_k} t_{j_k}^{-1}. $$
Its image under $\nu_P$ is non-zero if and only if this character is
trivial, which means that
$$|\{i_k\mid i_k=\ell\}|=|\{j_k\mid j_k=\ell\}| \quad \mbox{ for each} \quad
\ell\in \{1,\ldots, r\},$$
so that there exists a permutation $\sigma$ on $\{1,\ldots, n\}$ with
$j_k = i_{\sigma(k)}$ for $k =1,\ldots, n$. We thus obtain
\[
\nu\s{\al P.}.\big(A(P_{i_1}f_1,\ldots,P_{i_n}f_n;P_{j_1}g_1,\ldots,P_{j_n}g_n)\big)
=A(P_{i_1}f_1,\ldots,P_{i_n}f_n;P_{i_{\sigma(1)}}
g_1,\ldots,P_{i_{\sigma(n)}}g_n)\]
when it is nonzero.
Similar assertions hold for monomials in the $A$'s.

We now prove by induction on $n$ that the lemma holds for
$A \in \Spann\big(\bigcup_{j = 0}^n \cB_j\big)$.
If $n=1$, then we find for $f_1, g_1 \in \cH$:
$$ \nu\s{\al P.}.\big(A(f_1;g_1)\big)=
\nu\s{\al P.}.\Big(\sum_{i,\,j=1}^rA(P_{i}f_1;P_{j}g_1)\Big)=
\sum_{i=1}^rA(P_{i}f_1;P_{i}g_1) $$
leads to
$$ \big[\nu\s{\al P.}.\big(A(f_1;g_1)\big),\,a(f)\big]
=\sum_{i=1}^r\big[A(P_{i}f_1;P_{i}g_1),\,a(f)\big]
=-\sum_{i=1}^r \la P_{i}f_1,\,f\ra\,a(P_ig_1), $$
where we used
the identity
\begin{equation}\label{eq:car2}
[a(f)^*a(g), a(h)]
= a(f)^*a(g)a(h) - a(h) a(f)^* a(g)
= - \{a(f)^*,a(h)\} a(g)
= - \la f, h\ra  a(g).
\end{equation}
Thus we obtain
\begin{eqnarray*}
&& \left\|\big[\nu\s{\al P.}.\big(A(f_1;g_1)\big),\,a(f)\big]\right\|\leq
\sum_{i=1}^r\|P_if_1\|\,\|P_if\|\,\|P_ig_1\|
\leq\big(\max_{j}\|P_jf\|\big)\sum_{i=1}^r\|P_if_1\|\,\|P_ig_1\| \\[1mm]
&\leq& \big(\max_{j}\|P_jf\|\big)
\Big[\sum_{i = 1}^r \|P_if_1\|^2\Big]^{1/2}
\Big[\sum_{k = 1}^r \|P_k g_1\|^2\Big]^{1/2}
= \big(\max_{j}\|P_jf\|\big) \cdot \|f_1\|\,\|g_1\|,
\end{eqnarray*}
by the Cauchy--Schwarz inequality.
With the choice $C_A=\|f_1\|\,\|g_1\|,$
this proves our assertion for $A = A(f_1, g_1)$, and hence,
by the triangle inequality, our assertion for $n=1$ follows.

Next we assume that our assertion holds for each
element of $\Spann\Big(\bigcup_{j = 0}^k \cB_j\Big)$.
First we observe that
${\rm Span}\big(\al B._1\cup\cdots\cup\al B._{k+1}\big)$ is spanned by elements of the form
 $A=A_k+A_{k+1}$ where
$$A_k\in{\rm Span}\big(\al B._1\cup\cdots\cup\al B._{k}\big)
\quad \mbox{ and } \quad
A_{k+1}=A(f_1;g_1)\cdots A(f_{k+1};g_{k+1}). $$
Then our induction hypothesis implies that
\begin{eqnarray*}
\left\|\big[\nu\s{\al P.}.\big(A\big),\,a(f)\big]\right\|&\leq&
\left\|\big[\nu\s{\al P.}.\big(A_k\big),\,a(f)\big]\right\|+\left\|\big[\nu\s{\al P.}.\big(A_{k+1}\big),\,a(f)\big]\right\| \\[1mm]
&\leq& C_{A_k}\max_{\ell}\|P_\ell f\|+\left\|\big[\nu\s{\al P.}.\big(A_{k+1}\big),\,a(f)\big]\right\|,
\end{eqnarray*}
so that we only need to prove our assertion for $A_{k+1},$ i.e.
 we may assume that  $A =A(f_1;g_1)\cdots A(f_{k+1};g_{k+1})$.
Then
\begin{eqnarray*}
\nu\s{\al P.}.(A)&=&\nu\s{\al P.}.\Big(\sum_{\ell_1=1}^r\cdots\sum_{\ell_{2k+2}=1}^rA(P_{\ell_1}f_1;P_{\ell_{k+2}}g_1)\cdots
A(P_{\ell_{k+1}}f_{k+1};P_{\ell_{2k+2}}g_{k+1})\Big)  \\[1mm]
&=&\sum_{\ell_1=1}^r\cdots\sum_{\ell_{k+1}=1}^r \sum_{\sigma\in{\bf S}_{k+1}}
A(P_{\ell_1}f_1;P_{\ell_{\sigma({1})}}g_1)\cdots
A(P_{\ell_{k+1}}f_{k+1};P_{\ell_{\sigma(k+1)}}g_{k+1})
\end{eqnarray*}
where ${\bf S}_{k+1}$ denotes the permutation group of $\{1,\ldots,{k+1}\}.$
Observe that all the terms of this sum are in the image of
$\nu\s{\al P.}..$ Now
\begin{eqnarray*}
&&\big[\nu\s{\al P.}.(A),a(f)\big]
=\sum_{\ell_1=1}^r\cdots\sum_{\ell_{k+1}=1}^r \sum_{\sigma\in{\bf S}_{k+1}}
\Big\{\\[1mm]
&&\big[A(P_{\ell_1}f_1;P_{\ell_{\sigma({1})}}g_1),\,a(f)\big]\,A(P_{\ell_2}f_2;P_{\ell_{\sigma({2})}}g_2)\cdots
A(P_{\ell_{k+1}}f_{k+1};P_{\ell_{\sigma(k+1)}}g_{k+1})+\\[1mm]
&&\cdots+
A(P_{\ell_1}f_1;P_{\ell_{\sigma({1})}}g_1)\cdots
A(P_{\ell_{k}}f_{k};P_{\ell_{\sigma(k)}}g_{k})\,
\big[A(P_{\ell_{k+1}}f_{k+1};P_{\ell_{\sigma(k+1)}}g_{k+1}),\,a(f)\big]\Big\} \\[1mm]
&&=\sum_{\ell_1=1}^r\cdots\sum_{\ell_{k+1}=1}^r \sum_{\sigma\in{\bf S}_{k+1}}
\Big\{\\[1mm]
&&-(P_{\ell_1}f_1,f)a(P_{\ell_{\sigma({1})}}g_1)\,A(P_{\ell_2}f_2;P_{\ell_{\sigma({2})}}g_2)\cdots
A(P_{\ell_{k+1}}f_{k+1};P_{\ell_{\sigma(k+1)}}g_{k+1})-\\[1mm]
&&\cdots-
A(P_{\ell_1}f_1;P_{\ell_{\sigma({1})}}g_1)\cdots
A(P_{\ell_{k}}f_{k};P_{\ell_{\sigma(k)}}g_{k})\,
(P_{\ell_{k+1}}f_{k+1},f)a(P_{\ell_{\sigma({k+1})}}g_{k+1})\Big\}.
\end{eqnarray*}
We thus arrive at
\begin{eqnarray*}
&&\left\|\big[\nu\s{\al P.}.(A),a(f)\big]\right\|
\leq\sum_{\ell_1=1}^r\cdots\sum_{\ell_{k+1}=1}^r \sum_{\sigma\in{\bf S}_{k+1}}
\Big\{\\[1mm]
&&\|P_{\ell_1}f_1\|\,\|P_{\ell_1}f\|\,\|P_{\ell_{\sigma({1})}}g_1\|\,\|P_{\ell_2}f_2\|\,\|P_{\ell_{\sigma({2})}}g_2\|\cdots
\|P_{\ell_{k+1}}f_{k+1}\|\,\|P_{\ell_{\sigma(k+1)}}g_{k+1}\|+\\[1mm]
&&\cdots+
\|P_{\ell_1}f_1\|\,\|P_{\ell_{\sigma({1})}}g_1\|\cdots
\|P_{\ell_{k}}f_{k}\|\,\|P_{\ell_{\sigma(k)}}g_{k}\|
\|P_{\ell_{k+1}}f_{k+1}\|\,\|P_{\ell_{k+1}}f\|\,\|P_{\ell_{\sigma({k+1})}}g_{k+1}\|\,\Big\}\\[1mm]
&&\leq\Big(\max_{\ell}\|P_\ell f\|\Big)(k+1)
\sum_{\ell_1=1}^r\cdots\sum_{\ell_{k+1}=1}^r \sum_{\sigma\in{\bf S}_{k+1}}
\prod_{i = 1}^{k+1} \|P_{\ell_i} f_i\|
\prod_{j = 1}^{k+1} \|P_{\ell_{\sigma(j)}} g_j\|\\
&&= \Big(\max_{\ell}\|P_\ell f\|\Big)(k+1)
\sum_{\ell_1=1}^r\cdots\sum_{\ell_{k+1}=1}^r \sum_{\sigma\in{\bf S}_{k+1}}
\prod_{i = 1}^{k+1} \|P_{\ell_i} f_i\|
\prod_{j = 1}^{k+1} \|P_{\ell_j}g_{\sigma^{-1}(j)}\|\\
&&= \Big(\max_{\ell}\|P_\ell f\|\Big)(k+1)
\sum_{\sigma\in{\bf S}_{k+1}}\sum_{\ell_1=1}^r\cdots\sum_{\ell_{k+1}=1}^r
\prod_{i = 1}^{k+1} \|P_{\ell_i} f_i\|\|P_{\ell_i}g_{\sigma^{-1}(i)}\|\\
&&= \Big(\max_{\ell}\|P_\ell f\|\Big)(k+1)
\sum_{\sigma\in{\bf S}_{k+1}}
\prod_{i=1}^{k+1} \sum_{\ell_i=1}^r
\|P_{\ell_i} f_i\|\|P_{\ell_i}g_{\sigma^{-1}(i)}\|\\
&&\leq \Big(\max_{\ell}\|P_\ell f\|\Big)(k+1)
\sum_{\sigma\in{\bf S}_{k+1}}
\prod_{i=1}^{k+1} \|f_i\| \|g_{\sigma^{-1}(i)}\| \\
&&=  \Big(\max_{\ell}\|P_\ell f\|\Big)(k+1)(k+1)!
\prod_{i=1}^{k+1} \|f_i\| \|g_i\|,
\end{eqnarray*}
where we used the Cauchy--Schwarz inequality to obtain
\[
\sum_{\ell_i=1}^r\|P_{\ell_i}f_i\|\,\|P_{\ell_i}g_{\sigma^{-1}(i)}\|
\leq\Big[\sum_{i = 1}^r \|P_{\ell_i}f_i\|^2 \Big]^{1/2}
\Big[\sum_{i = 1}^r \|P_{\ell_i} g_{\sigma^{-1}(i)}\|^2\Big]^{1/2}
= \|f_i\|\,\|g_{\sigma^{-1}(i)}\|.\]
Observe that
$C_A := (k+1)(k+1)! \prod_{i=1}^{k+1} \|f_i\| \|g_i\|$
does not depend on $\cP$, so that this completes our induction.
\end{prf}
We are now ready to prove our first main result.
\begin{teo}   \label{TrivInv1}
Let $\cH$ be complex Hilbert space and let
$\cN \subeq B(\cH)$ be a non-atomic commutative von Neumann algebra.
Then $\Car(\cH)^{\U(\cN)} = \C \1$.
\end{teo}

\begin{prf}  If an $A\in X_0 := {\rm Span}\Big(\bigcup\limits_{i=0}^\infty
\al B._i\Big)$ is $\U(\cN)$-invariant, then
$\nu\s{\al P.}.(A)=A$ for all partitions $\cP\subset\cN.$ Hence the preceding lemma
leads to
$$ \big\|[A,\,a(f)]\big\|\leq C_A\max_{\ell}\|P_\ell f\|\quad\hbox{for all}\quad
\; f \in \cH,\;\cP\subset\cN,$$
where $C_A$ does not depend on $\cP$ or $f.$
Since $\cN$ is nonatomic, Proposition~\ref{NonatomP} implies that
$\max_{\ell}\|P_\ell f\|$ can be made arbitrarily small.
We conclude that $A$ commutes with all $a(f),$ $f\in\al H.,$
hence with $\Car(\cH)$. But the center of $\Car(\cH)$ is $\C \1$,
which leads to
\[
{\rm Span}\Big(\bigcup\limits_{i=0}^\infty\al B._i\Big)\cap
\Car(\cH)^{\U(\cN)}=\C\1.
\]
Recall that $\Gicar(\cH) = \Car(\cH)^{\T\1}$
is the closure of $X_0$. Let
$A \in \Gicar(\cH)^{\U(\cN)}=\Car(\cH)^{\U(\cN)}$, then for each $\varepsilon>0$ there exists an
$A_0\in X_0$ such that ${\|A-A_0\|}<\varepsilon.$
For a unit vector $f \in \cH$
we  choose the finite partition $\cP$ in such a way that
$C_{A_0} \max_{\ell} \|P_\ell f\| < \eps$.
Then we obtain
\begin{eqnarray*}
\left\|\big[A,a(f)\big]\right\|&=&\left\|\big[\nu\s{\al P.}.(A),a(f)\big]\right\| \leq
\left\|\big[\nu\s{\al P.}.(A-A_0),a(f)\big]\right\| +\left\|\big[\nu\s{\al P.}.(A_0),a(f)\big]\right\| \\[1mm]
&\leq&2\|f\|\,\|A-A_0\|+C_{A_0}\max_{\ell}\|P_\ell f\|\leq
2\varepsilon+\eps = 3\eps.
\end{eqnarray*}
Since $\eps > 0$ was arbitrary, we obtain
$A \in Z(\Car(\cH)) = \C \1$, and this completes the proof.
\end{prf}

\begin{rem} If the von Neumann algebra $\cN$ is not non-atomic,
then we obtain for each minimal non-zero projection $P \in P_\cN$
a decomposition
$$\U(\cN)
\cong e^{i\R P} \times \U((\1-P)\cN)
\cong \T \times \U((\1-P)\cN),  $$
so that
$$ \C \1 \not= \Gicar(P(\cH)) \subeq \Car(\cH)^{\U(\cN)}. $$
Therefore the assumption of $\cN$ being non-atomic in the
preceding theorem is necessary.
\end{rem}

For maximal commutative subalgebras, the preceding theorem
could also be obtained from  the results of Wolfe in \cite{Wol}.
However his arguments are very indirect and difficult.
We think that our proof is much more transparent and direct.
\begin{cor}
\label{denseAlg}
If $\cA \subeq B(\cH)$ is a $C^*$-subalgebra, then
$\Car(\cH)^{\U(\cA)}= \Car(\cH)^{\U(\cA'')},$ and if
$\cA''\subeq B(\cH)$ contains a nonatomic commutative von Neumann algebra,
then
$\Car(\cH)^{\U(\cA)}= \C\1$.
\end{cor}

\begin{prf}
The action of $\U(\cH)$ on $\Car(\cH)$ is continuous. Since
$\U(\cA)$ is strongly dense in
$\U(\cA'')$ by Kaplansky's Density Theorem (\cite[Cor.~5.3.7]{KR1}),
it follows that $\Car(\cH)^{\U(\cA)}= \Car(\cH)^{\U(\cA'')}.$
Let $\cN\subset\cA''$ be a nonatomic commutative von Neumann subalgebra,
then by $\U(\cN)\subset \U(\cA'')$ we get
$\C\1=\Car(\cH)^{\U(\cN)}\supseteq\Car(\cH)^{\U(\cA'')}=\Car(\cH)^{\U(\cA)}.$
\end{prf}
\begin{rem} \label{rem:tens}
To prepare for our second main result, we need to recall some facts about tensor products
(cf. \cite[Sect.~IV.2, p188]{Tak}).
Let $X$ and $Y$ be Banach spaces and let $X'$, resp., $Y'$ be their
topological duals. We have an identification of  the algebraic
tensor product $X \otimes Y$ with a subspace of $B(X',Y)$
by the linear injection $\Phi:X \otimes Y\to B(X',Y)$
given  by ${\Phi(x\otimes y)(f)}:=f(x)y$ for $f\in X',$ and we have a similar identification map
$\Psi:X \otimes Y\to B(Y',X)$ by ${\Psi(x\otimes y)(f)}:=f(y)x$ for $f\in Y'.$
 The map $\Phi:X \otimes Y\to B(X',Y)$ is an isometry w.r.t. the
 minimal cross-norm $\lambda,$ hence extends as an isometry to the completion,
 denoted by  $X \otimes_\lambda Y$ (cf. \cite[Prop.~IV.2.1, p189]{Tak}).
Explicitly $\|\cdot\|_\lambda$ is given by
\[
\Big\|\sum_{i=1}^nx_i\otimes y_i\Big\|_\lambda:=
\sup\Big\{\Big|\sum_{i=1}^nf(x_i)\, g(y_i)\Big|\,\mid\,
f\in X',\, \|f\|\leq1;\; g\in Y',\|g\|\leq1\Big\}.
\]
It is easy to verify that $\otimes_\lambda$ is a bifunctor
on the category of Banach spaces. This implies in particular
that, if  $X = X_1 \oplus X_2$ is a direct sum of two closed subspaces
and $P_j \: X \to X_j$ are the corresponding projections,
then we obtain a topological
isomorphism
$$ X \otimes_\lambda Y \to
(X_1 \otimes_\lambda Y) \oplus (X_2 \otimes_\lambda Y), \quad
\Psi(x \otimes y) \mapsto
(P_1 \circ \Psi(x \otimes y)) \oplus
(P_2 \circ \Psi(x \otimes y)) $$
subordinated to the decomposition
$B(Y',X) \cong B(Y',X_1) \oplus B(Y',X_2)$.
From that we further conclude that
\begin{eqnarray}
\Phi\big(X_1 \otimes_\lambda Y\big)&=&\{ \phi \in \Phi\big(X \otimes_\lambda Y\big) \subeq B(X',Y) \,\mid\,
X_1^\bot \subeq \ker \phi\}  \nonumber\\
\label{X1char}
\Psi\big(X_1 \otimes_\lambda Y\big)
&=&\{ \psi \in \Psi(X \otimes_\lambda Y) \subeq B(Y',X) \,\mid\,
 \im(\psi) \subeq X_1\}.
\end{eqnarray}
This observation is particularly useful if
$G \subeq \GL(X)$ is  a group of isometries and
$X_1 = X^G$ is the subspace of $G$-fixed elements.
Then $G \otimes \id_Y$ is a group of isometries of $X \otimes_\lambda Y$
for which ${\Psi\big((g\otimes \id_Y)(x\otimes y)\big)(f)}:=f(y)\,g.x$ for $f\in Y',$
i.e. $\Psi$ is $G$-equivariant. In particular
${\Psi\big((X \otimes_\lambda Y)^{G \otimes \id_Y}\big)(f)}\subseteq X^G$ for $f\in Y'.$
Thus, since $X = X^G \oplus X_2,$ we get from (\ref{X1char}) that
\begin{equation}
  \label{eq:fixrel}
(X \otimes_\lambda Y)^{G \otimes \id_Y} = X^G \otimes_\lambda Y.
\end{equation}
Note that the minimal C*-cross norm for nonabelian C*-algebras does not coincide with
the minimal Banach cross norm $\lambda$
(cf. \cite[Theorem~IV.4.14, p189]{Tak}).
\end{rem}

\begin{pro}
\label{TrivInv2}
Given a decomposition
$\al H.=\al K.\oplus\al L.$
of a complex Hilbert space,
let $\al U.\subset \U(\cK)$
be a group of unitaries with
$\Car(\cK)^\cU =\C\1$.
Then $\Car(\cH)^\cU = \Car(\cL),$
considered as a subalgebra of $\Car(\cH)$,
where we identified $\al U.\subset \U(\cK)$
with $\al U.\oplus\1\subset \U(\cH).$
\end{pro}

\begin{beweis}
The operator $V:=P\s{\al K.}.-P\s{\al L.}.$ on $\cH$ is unitary
and the associated automorphism
of $\Car(\cH)$ satisfies
$\alpha_V\restriction\Car(\al K.)=\id=\alpha_V^2,$ and
$\alpha_V\restriction\Car(\al L.)$ induces the canonical grading
for which the generators $a(f)$ are odd.
In particular, $P^{\al L.}_{\rm even}:=\hlf(\id+\alpha_V)$
projects $\Car(\al H.)$
onto
$${C^*\big(\Car(\cK)\cup\Car(\cL)_{\rm even}\big)}\cong
{\Car(\al K.)\otimes\Car(\al L.)_{\rm even}}$$
(\cite[Exer.~XIV.1.5b,p.~94]{Tak3}),
 and
$P^{\al L.}_{\rm odd}:=\hlf(\id-\alpha_V)$ projects $\Car(\al H.)$
onto the subspace ${\overline{\rm span}\big(\Car(\al K.)\cdot\Car(\al L.)_{\rm odd}\big)}.$
Since ${[\alpha_U,\alpha_V]}=0$ for all $U\in\al U.(\al H.)$ preserving
$\cL$, the relation $A\in\Car(\al H.)^{\al U.}$ is equivalent to
$P^{\al L.}_{\rm even}A\in\Car(\al H.)^{\al U.}\ni P^{\al L.}_{\rm odd}A $.
Thus, for the rest of this proof, we may assume that
either $A\in P^{\al L.}_{\rm even}\Car(\al H.)$ or
$A\in P^{\al L.}_{\rm odd}\Car(\al H.)\,.$

First, let
\[
A\in\Car(\al H.)^{\al U.}\cap P^{\al L.}_{\rm even}\Car(\al H.)
\subset\Car(\al K.)\otimes\Car(\al L.)_{\rm even}
\subseteq\Car(\al K.)\otimes_\lambda \Car(\al L.)_{\rm even}.
\]
Here we use the fact that the C*-tensor product of $C^*$-algebras
is defined by a cross norm which dominates the minimal Banach cross norm $\lambda$, which
leads to the inclusion on the right (cf.\ Remark~\ref{rem:tens}).
Now observe that $\C\1=\Car(\cK)^\cU$ is complemented, in fact a projection onto $\C\1$ is given by
$P_f(A):=f(A)\1$ for any continuous functional $f$ with $f(\1)=1,$  in which case the complementary subspace is
$\ker(f).$ Thus
\eqref{eq:fixrel} above implies that
$$ (\Car(\al K.)\otimes_\lambda \Car(\al L.)_{\rm even})^\cU
=  \Car(\cK)^\cU \otimes_\lambda \Car(\cL)_{\rm even}
=  \1 \otimes_\lambda \Car(\cL)_{\rm even}, $$
which implies that $A \in \Car(\cL)_{\rm even}$.

Now let $A\in P^{\al L.}_{\rm odd}\Car(\al H.)$ be a $\cU$-invariant
element. We observe that for every unit vector $f \in \cL$,
the element $u := a(f) + a(f)^*$ is hermitian and satisfies
$$ u^2 = a(f)a(f)^* + a(f)^* a(f) = \{a(f), a(f)^*\} = \la f, f \ra \1= \1,$$
so that it is unitary. Clearly, $u \in \Car(\cL)_{\rm odd}$. Then
$A u \in P^{\al L.}_{\rm even}\Car(\al H.)
$ is also
$\cU$-invariant, hence contained in $\Car(\cL)_{\rm even}$
by the preceding argument.
This leads to $A = Auu \in \Car(\cL)_{\rm odd}$, so that our proof
is complete.
\end{beweis}

\section{Automorphism Groups Encoding Locality}

To build the quantum fields, we need first to add extra structure to the
classical fields, which we now list.
Let $M$ be a $k$-dimension $\sigma$-compact
smooth manifold and let $G \subeq \U_n(\C)$ be a closed subgroup.
Further, let $\q_E:E\to M$ be a complex vector bundle which
is a G-bundle, i.e. its typical fibre is $\C^n$ on which $G$
acts by the defining matrices for $\U_n(\C)\supset G,$ and
$E$ has an atlas of local trivializations for which the
transition functions take their values in $G$ (the set of trivializations is called a G-structure).
It is known that the property of $E$ being a G-bundle
 is equivalent to requiring $E$ to be
associated to a $G$-principal bundle $\q_P:P \to M$ with respect to the
identical representation of $G$ on $\C^n$
(cf.~\cite[Corr.~37.13]{KM97} and \cite[p.~368]{CB}).
This means that
$E = (P \times \C^n)/G$, where $G$ acts on $P \times\C^n$ by
$g.(p,\bv) = (pg^{-1}, g\bv)$ and $\q_E\big([p,\bv]\big):= \q_P(p)\in M$
where we write $[p,\bv]$ for the elements
of $E$, i.e. the $G$-orbit of $(p,\bv)$ in $P \times \C^n.$

We obtain from
the norm on $\C^n$ a function
$$|\cdot|:E \to[0,\infty)\quad\hbox{by}\quad
\big|[p,\bv]\big|= |\bv| =(\overline{\b v.}\cdot\b v.)^{1/2}. $$
This is well--defined because $G$ acts as unitaries on $\C^n\,.$
Let $\mu$ be a locally
Lebesgue measure on $M$ (this could be obtained from a
nowhere vanishing smooth $k\hbox{--form}$  on $M$ if we are concerned with smoothness)
and consider the $L^2\hbox{--sections}$ of $E\,,$ i.e. $c:M\to E$
smooth such that $\q_E\circ c={\rm id}_M$ and
$$\|c\|^2:=\int_M |c(x)|^2\,d\mu(x)<\infty\,.$$
Let $L^2(E)$ denote the  $L^2\hbox{--completion}$ of the
space of smooth $L^2\hbox{--sections}$ w.r.t. this norm.

Now the $C^\infty(M)\hbox{--module}$ action on $\Gamma(E)$ need to be restricted
to the bounded smooth functions  $C_b^\infty(M)$ to obtain an action on
$L^2(E)$ by bounded operators. If we complete $C_b^\infty(M)$
w.r.t. the strong operator topology we get an action of
$L^\infty(M,\mu)$ on $L^2(E)\,.$
This is because the subalgebra $C_c^\infty(M)$ separates all the points of $M$
and on each point is nonzero, hence by the Stone--Weierstrass theorem it is C*-norm dense
in $C_0(M),$ and the von
Neumann algebra generated by $C_0(M)$ in $B\big(L^2(E)\big)$
is $L^\infty(M,\mu).$
Note that this module action of $L^\infty(M,\mu)$ on $L^2(E)$ encodes locality of the classical
fields in a particularly simple way, e.g. restriction to a Borel subset $W\subset M$ with $\mu(W)\not=0,$
is just done by multiplication of the characteristic function $\chi\s W.\in L^\infty(M,\mu).$
We denote this submodule by $L^2(E\restriction W):=\chi\s W.L^2(E).$
It is characterized by
\[
L^2(E\restriction W)=\big\{c\in L^2(E)\,\mid\,f\cdot c=0\;\;
\forall\,f\in\chi\s W^c. L^\infty(M,\mu)=L^\infty(W^c,\mu)\big\}\,.
\]
Moreover, $L^\infty(M,\mu)$ is nonatomic, as
$\mu$ is locally  Lebesgue. Henceforth we omit $\mu$ from the notation $L^\infty(M,\mu).$
Thus, from the original $C^\infty(M)\hbox{--module}$ action on $\Gamma(E)$ we have obtained an action of the
commutative  nonatomic von Neumann algebra  $L^\infty(M)$ on $L^2(E)\,,$
and it encodes locality information of the classical fields.
Below we will use its unitary group to define
automorphisms of the CAR--algebra of $L^2(E)$.

To quantize the matter fields, we consider the $C^*$-algebra
$\Car(L^2(E)),$ on which we have
the usual action $\alpha:\al U.(\al H.)\to\aut{\rm CAR}(\al H.),$ $\al H.:=L^2(E),$
given by ${\alpha\s U.\big(a(f)\big)}:=a(Uf),$ $f\in\al H..$
Since for classical gauge theory, the $C^\infty(M)\hbox{--module}$ action
on the matter fields is crucial,
we will be particularly concerned with the restriction of $\alpha$ to
$\U\big(L_\C^\infty(M)\big)\subset \U(\cH).$

In the quantum situation, locality is specified by the algebras
$$\al A.(W):={C^*\{a(c)\mid c\in L^2(E\restriction W)\}}\cong
{\rm CAR}(L^2(E\restriction W))$$
for any relatively compact open set $W\subset M.$ This collection of algebras
satisfies precisely the conditions (L1) and (L2) above, where we take
$\al F.=\CAR.$
We now show that each $\al A.(W)$ is in fact the fixed point algebra in $\CAR$ of
the automorphism group $\alpha\s \U(L_\C^\infty(W^c)).,$ hence
$\alpha\restriction \U\big(L_\C^\infty(M)\big)$ already  contains the
locality information in this quantum context. The natural identification of
$\U(L_\C^\infty(W^c))$
with the unitaries in $\U\big( L_\C^\infty(M)\big)$
which are $1$ on $W,$ i.e. with $\chi_W\oplus \U(L_\C^\infty(W^c)),$ was used.
In view of the preparations from the preceding sections,
we can now prove our main result.

\begin{teo} With respect to the action
$\alpha:\U\big( L_\C^\infty(M)\big)\to\aut\CAR$ from above, for
each relatively compact $W\subset M$,
the subalgebra $\al A.(W)$ is  the fixed point algebra of
the subgroup $\U(L_\C^\infty(W^c))$ of $\U(L_\C^\infty(M))$,
where $W^c$ denotes the complement of $W.$
\end{teo}

\begin{beweis} The von Neumann algebra
$L^\infty_\C(W^c)$ is nonatomic since  $\mu$ is locally Lebesgue.
Thus $L^\infty_\C(W^c)$ satisfies the hypotheses of
Theorem~\ref{TrivInv1}. Hence
the fixed point algebra of
the group of automorphisms $\alpha\s \U(L_\C^\infty(W^c)).$ acting on
${\rm CAR}(L^2(E\restriction W^c))$ is just the constants.
Since $L^2(E)=L^2(E\restriction W)\oplus
L^2(E\restriction W^c)$,  it now follows from Proposition~\ref{TrivInv2} that
$\al A.(W)$ is  the fixed point algebra of
 $\alpha\s \U(L_\C^\infty(W^c)).\subset\aut\CAR$ as claimed.
\end{beweis}

\section{Invariance Under Local Gauge Transformations}

We can now use our result above to prove a well-known ``folk theorem,'' stating that the only invariant elements
under the group of local gauge transformations
of the CAR--algebra are the constants.  We will need the very mild condition that $\T\1\subseteq G.$
This condition guarantees that   $\alpha\s \U(C_b^\infty(M)).$ is contained in the action of the local
gauge group on the CAR--algebra, and we will need only enough detail of the gauge action on $\CAR$
to verify this.

An intrinsic definition of  the gauge group $\gau E$ is as the group of those smooth bundle automorphisms
$\gamma\in\Aut(E)$  which induce the identity
on the base manifold $M$,
i.e. $\q_E \circ \gamma = \q_E,$
and which preserves the G-structure, i.e. the union of the G-structure  and
its composition with $\gamma$ is a G-structure. However, the (equivalent) customary definition
of $\gau E$ is via the property that $E= (P \times \C^n)/G$ is an associated bundle to $P.$
Briefly, one defines $\gau E$ as those $\gamma\in\Aut(E)$ of the form
$\gamma[p,\bv] = [p, f(p)\bv]$, where $f \: P \to G$ is a smooth function
satisfying $f(p.g) = g^{-1} f(p)g$ for all $p\in P,\;g\in G$
(cf. \cite[Thm.~3.2.2]{Bl81} and \cite[Comment~4, p.~239]{Is99}).
 If $f$ has values in the
center of $G$, then it defines a function $\oline f:M\to Z(G)$ by
$f=\oline f\circ\q_P,$ and $\gamma$ commutes with $\gau E.$

To obtain a unitary action on $L^2(E)$ from the action of $\gau E$ on $E,$
observe that ${\big|\big(\gamma\cdot c\big)(x)\big|}=\big|c(x)\big|$
for $\gamma\in\gau E$ and $c\in L^2(E),$
which leads us to define
\[
\big(V_\gamma c\big)(x):=\big(\gamma\cdot c\big)(x)
:=\gamma\big(c(x)\big)\,.
\]
If $f$ has values in the
center of $G$, then  $\big(V_\gamma c\big)(x)=
\oline f(x)\,c(x).$ In particular, if $\T\1\subseteq G\subset \U(\C^n),$
then $V\s\gau E.$ contains the module action of $\U(C_b^\infty(M))$
on $L^2(E).$
Using the action $V:\gau E\to\U(L^2(E)),$ we obtain the usual automorphic action
\[
\kappa:\gau E\to\aut\CAR\quad\hbox{by}\quad
\kappa_\gamma\big(a(f)\big):=a(V_\gamma f)\quad\mbox{ for } \quad
\gamma\in \gau E,\; f\in L^2(E). \]
Note that $\kappa(\gau E)$ commutes with
$\alpha\big(\U\big( L_\C^\infty(M)\big)\big)$
since the multiplication with smooth functions
commutes with the action $V$ of $\gau E$ on $\Gamma(E)$.
\begin{teo} \label{thm:folk}
Given the action $\kappa:\gau E\to\aut\CAR$ above, then \chop
$\CAR^{\gau E}=\C\1$ if either:
\begin{itemize}
\item[\rm(i)] $\T\1\subseteq G,$ or
\item[\rm(ii)] in the Fock representation $\pi_\cF$ we have that
$\Gamma(V(\gau E))''\supset\Gamma(\U(L^\infty_{\C}(M)),$ where $\Gamma$ denotes the
second quantization on Fock space of unitaries on $L^2(E).$
\end{itemize}
\end{teo}
\begin{beweis}
(i) If $\T\1\subset G,$ then
 $\alpha\s \U(C_b^\infty(M)).\subset\kappa(\Gau
E),$ and hence any gauge invariant element
is invariant w.r.t. $\alpha\s \U(C_b^\infty(M))..$
Observe  that $C_b^\infty(M)''=C_b(M)''=L^\infty(M),$
hence for any selfadjoint $H\in L^\infty(M),$ there is a net
$\{H_\nu\}\subset C_b^\infty(M)$ of selfadjoint elements such that
$H_\nu\to H$ in the strong operator topology, and with $\|H_\nu\|\leq\|H\|$ for
all $\nu$ (cf.
\cite[Thm.~5.3.5, p.~329]{KR1}).
However, any unitary $U\in L^\infty(M)$ is of the form
$U=\exp(iH)$ for some positive $H\in L^\infty(M)$
(cf. \cite[Thm.~5.2.5, p.~313]{KR1}).
Since  the map $t\to\exp(it)$ is continuous, we have
\[
\U(C_b^\infty(M))\ni \exp(iH_\nu)\to\exp(iH)=U\in L^\infty(M)\,
\]
in the strong operator topology (cf. \cite[Prop.~5.3.2, p327]{KR1}).
Hence by continuity of the action $\alpha,$
we have by Theorem~\ref{TrivInv1} that
\[
\CAR^{\gau E}\subseteq\CAR^{\U(C_b^\infty(M))}= \CAR^{\U(L^\infty(M))}=\C\1\,.
\]
(ii)  If $\Gamma(V(\gau E))''\supseteq\Gamma(\U(L^\infty_{\C}(M))),$ then
$\Gamma(V(\gau E))'\subseteq\Gamma(\U(L^\infty_{\C}(M)))',$ and as
\[
\Gamma(\U(L^\infty_{\C}(M)))'\cap\pi_\cF\big(\CAR\big)=\pi_\cF\Big(\CAR^{\U(L^\infty_{\C}(M))}\Big)
=\C\1,
\]
it follows that
\[
\Gamma(V(\gau E))'\cap\pi_\cF\big(\CAR\big)=\pi_\cF\big(\CAR^{\gau E}\big)
\subeq \C \1.
\]
Since $\pi_\cF$ is faithful (as $\CAR$ is simple),
we obtain that $ \CAR^{\gau E}=\C\1\,.$
\end{beweis}
In a full gauge theory, electromagnetism will be
included, hence
$G=\nu\big(\T\times H\big)$ where $H$ is a compact connected Lie group,
and  $\nu:\T\times H\to \U(\C^n)$ is a homomorphism which takes $(\T,e)$ to $\T\1$
(cf.~\cite[p.~118]{CG07}).
Thus the assumption of the inclusion $\T\1\subseteq G$ is not physically unreasonable.

Theorem~\ref{thm:folk} seems to be a well--known ``folk theorem,''
and the usual strategy for finding gauge invariant elements
in an appropriate representation $\pi:\CAR\to\al B.(\al H.)$
is to select them from the currents (generators of the unitary
one--parameter groups inplementing automorphisms of the $\CAR$).
Of course, if one enlarges $\CAR$ by additional elements which are gauge invariant,
 the action of $\U(L^\infty(M))$ will not be able to
select the local algebras directly.

\section{Localizing Maps}

There is more useful information
in the automorphic action $\alpha:\U\big( L_\C^\infty(M)\big)\to\aut\CAR,$
other than the net of local algebras.
For example, we can obtain from it a set of conditional expectations which
extend the restriction
maps $L^2(E) \to L^2(E\restriction S),
c\to c\restriction S$   from the fields to the algebra $\CAR$ (for $S\subset M$ open and relatively compact).
Note that the map $a(f)\to a(f\restriction S)$ cannot extend to $\CAR$
as a *-homomorphism, since this will violate the canonical anticommutation relations.

Recall that a map $\nu:\CAR\to\CAR$ is a conditional expectation if
it is a positive map such that $\nu(\1)=\1$ and
\[
\nu\left(A\nu(B)\right)=\nu\left(\nu(A)B\right)=\nu(A)\nu(B)\quad
\forall\;A, B\in\CAR.
\]
A projection on $\CAR$ is a contractive linear map which is idempotent.
In \cite[Thm.~III.3.5]{Tak},
it is shown  that all projections which preserve the identity are
in fact conditional expectations. The maps $\nu_{\al P.}:\Car(\cH)\to\Car(\cH)^T$
which occurred in the proof of Lemma~\ref{lem:finpro}
are typical examples.

\begin{teo}
\label{rhorestrict}
Let $S\subset M$ be a relatively compact set. Then there is a
conditional expectation ${\nu_S:\CAR}\to\al A.(S)\subset\CAR,$ which
 satisfies $\nu_S(\alpha_h(A))=\nu_S(A)$ for all
$h\in \U(L_\C^\infty(S^c)), $  and
\begin{equation}
  \label{eq:nuS}
\nu_S\big(A(f_1,\ldots,f_n;g_1,\ldots,g_m)\big)=A\big(p_Sf_1,\ldots,p_Sf_n;p_Sg_1,\ldots,p_Sg_m\big)
\quad\mbox { for} \quad f_i,g_j\in L^2(E),
\end{equation}
where
$A(f_1,\ldots,,f_n;g_1,\ldots,g_m):=a^*(f_1)\cdots a^*(f_n)a(g_1)\cdots a(g_m)$ denotes a normal ordered monomial,
and $p_S$ is the projection of $L^2(E)$ onto the subspace $L^2\big(E\restriction S\big)$.
\end{teo}
\begin{beweis}
We can build $\nu_S$ from the conditional expectations $\nu_{\al P.}$
from Lemma~\ref{lem:finpro} (\cite[Prop.~3.11]{Wol}), but
it is easiest just to define it explicitly. Recall first  that, given
$R\in\cl B.(L^2(E))$ with $\|R\|\leq 1$,
 we can define a positive map $\alpha_R:\CAR\to\CAR,$ which takes
each normal ordered monomial $a^*(f_1)\ldots a^*(f_n)a(g_1)\ldots a(g_m)$
to $a^*(Rf_1)\ldots a^*(Rf_n)a(Rg_1)\ldots a(Rg_m)$  and the identity
to itself (cf.~\cite[Prop.~2.1]{HuKa}).
With $R=p_S$ and $\nu_S:=\alpha_{p_S},$ we then obtain
\eqref{eq:nuS}.
It is uniquely determined by what it does on the monomials $A(\cdots)$ since
$\CAR$ is topologically spanned by these.
Its equivariance w.r.t.\ $\U(L_\C^\infty(S^c)$ is obvious, as well as the fact that it is idempotent,
hence a conditional expectation \cite[Thm.~III.3.5]{Tak}.
\end{beweis}

Another useful approach to the restriction maps is
as follows.
For any Borel set $S\subset M$,  let
$\al N._{S}\subset\CAR$ be the closed left ideal of $\CAR$ generated by the set
${\{a(c)\,\mid\,c\in L^2(E\restriction S)\}},$ and denote the generating hereditary subalgebra
by $\al D._S:=\al N._S\cap\al N._S^*.$ Note that $\al N._{S}$ is proper
since it annihilates the vacuum in the Fock representation.
If $S$ is relatively compact, then $\al N._{S^c}$
is nonzero for $M$ noncompact.
Now each closed left ideal $\al J.$ of $\CAR$ has a unique associated open projection
in the universal von~Neumann algebra
$P\in\CAR''$ characterized by $\al J.={\CAR\cap\CAR''}P$ (cf.~\cite[Prop.~3.11.9, 3.11.10, Thm.~3.10.7]{Ped}, \cite{Ak1}), and then the hereditary $C^*$-subalgebra is
$$\al J.\cap\al J.^*={\CAR\cap P\,\CAR''}P.$$
Hence for any state $\omega$ we have $\omega(P)={\|\omega\restriction(\al J.\cap\al J.^*)\|}.$

\begin{pro}
Let  $S$ be relatively compact  and let
 $P_{S^c}$ be the open projection of $\al N._{S^c},$
and denote its  complementary closed projection by $\overline{P}_S:=\1- P_{S^c}.$
Define a map
\[
\wt\nu_S:\CAR\to{\rm C}^*\left(\{P_{S^c}\}\cup\CAR\right)\quad\hbox{by}\quad
\wt\nu_S(A):=\overline{P}_SA\overline{P}_S\,.
\]
Then we have that
\[
\wt\nu_S(A)=\overline{P}_S\nu_S(A)\overline{P}_S\quad
\hbox{for all}\quad A\in\CAR.
\]
 Moreover $\wt\nu_S$ is an isomorphism on $\al A.(S),$ i.e.
$\nu_S\left(\CAR\right)=\al A.(S)\cong {\overline{P}_S\al A.(S)\overline{P}_S}=\wt\nu_S\left(\CAR\right),$
and furthermore
\[
\al N._{S^c}=\left\{A\in\CAR\;\mid\;\nu_S(A^*A)=0\right\}.
\]
\end{pro}
\begin{beweis}
Since $P_{S^c}$ acts as a right identity for the elements of $\al N._{S^c}$ we have that\chop
$a(c)(\1-P_{S^c})=0$ whenever $c\in L^2(E\restriction S^c).$ Hence,
 for all $c\in L^2(E)$, we have
$$ a(c)(\1-P_{S^c}) = \big(a(p_Sc)+a((\1-p_S)c)\big)
(\1-P_{S^c})=a(p_Sc)(\1-P_{S^c}),$$
and hence
$$ (\1-P_{S^c})a(c)^*=(\1-P_{S^c})a(p_Sc)^*.$$
Given any normal ordered monomial
$$ A(c_1,\ldots,,c_n;d_m,\ldots,d_1) :=
a(c_1)^*\cdots a(c_n)^*a(d_m)\cdots a(d_1), $$
we can permute the $a(c_i)^*$ amongst themselves, and the $a(d_j)$ factors amongst themselves
using the CAR--relations, acquiring only $\pm$ factors in the process.
 Thus using such permutations to get terms adjacent to $\1-P_{S^c}$,
we obtain:
\begin{eqnarray*}
&&\wt\nu_S(A(c_1,\ldots,,c_n;d_m,\ldots,d_1))
= (\1-P_{S^c}) A(c_1,\ldots,,c_n;d_m,\ldots,d_1)(\1-P_{S^c})\\[1mm]
&&\quad= (\1-P_{S^c})A(p_Sc_1,\ldots,,p_Sc_n;\,p_Sd_m,\ldots,p_Sd_1)(\1-P_{S^c})\in
(\1-P_{S^c})\al A.(S)(\1-P_{S^c}).
\end{eqnarray*}
Thus the positive map $\wt\nu_S$ will take all normal ordered
polynomials in the fields to the same normal ordered polynomials of the fields
restricted to $S,$ conjugated by $\1-P_{S^c}=\overline{P}_S.$ Thus $\wt\nu_S(A)=\overline{P}_S\nu_S(A)\overline{P}_S$
for all $A\in\CAR.$
In particular, the range of $\wt\nu_S$ is
${\overline{P}_S\al A.(S)\overline{P}_S}.$

Next, we show that on $\al A.(S)$ the map
$\wt\nu$ is in fact an isomorphism. It suffices to show that
$P_{S^c}$ commutes with $\al A.(S),$ since then $\wt\nu_S$ is a *--homomorphism, which
is an isomorphism since $\al A.(S)={\rm CAR}(L^2(E\restriction S))$ is simple.
Recall that  $P_{S^c}$ is the complementary projection of the projection onto the subspace
annihilated by ${\{a(c)\,\mid\,c\in L^2(E\restriction S^c)\}}$ in the universal representation space.
The generating elements ${\{a(c)\,\mid\,c\in L^2(E\restriction S)\}}$ of $\al A.(S)$
and their adjoints all anticommute with  ${\{a(c)\,\mid\,c\in L^2(E\restriction S^c)\}},$
hence will preserve the subspace  annihilated by ${\{a(c)\,\mid\,c\in L^2(E\restriction S^c)\}}.$
Thus ${\{a(c)\,\mid\,c\in L^2(E\restriction S)\}}$ (as well
as $\al A.(S))$ will commute with the projection
onto this subspace, hence with $P_{S^c}.$ As $\wt\nu_S$ is thus an isomorphism on $\al A.(S)$
it follows for its range that
${(\1-P_{S^c})\al A.(S)(\1-P_{S^c})}\cong\al A.(S).$

 To see that
 $\al N._{S^c}=\left\{A\in\CAR\;\mid\;\nu_S(A^*A)=0\right\},$
  note that by the previous isomorphism, $0=\nu_S(A^*A)$ if and only if
$0=\wt\nu_S(A^*A)={(\1-P_{S^c})A^*A(\1-P_{S^c})}$
  if and only if $A=AP_{S^c},$ which characterizes $\al N._{S^c}.$
\end{beweis}
Thus we have shown that the classical module action of $C^\infty(M)$ on the
matter fields defines
an automorphism group on $\CAR$ which provides
the analogous localization to what this module action
does in the classical picture. We also identified the quantum analogues of restriction maps,
and obtained a proof that the $\CAR$ has only trivial gauge invariant elements.

\section*{Acknowledgements}

HG wishes to thank the DAAD for generously funding his visit to Germany in  2009, as well as the Mathematics Department of the Darmstadt University
of Technology for supporting his visit to Darmstadt.

\end{document}